\documentclass[10pt,conference]{IEEEtran}
\usepackage{cite,graphicx,booktabs,array,balance,multirow}
\usepackage[table]{xcolor}
\usepackage[hidelinks]{hyperref}
\usepackage{xurl}
\usepackage{tikz}
\usepackage{pgfplots}
\usepgfplotslibrary{groupplots}
\pgfplotsset{compat=1.18}
\usepackage{listings}
\usetikzlibrary{positioning,arrows.meta,fit,calc}

\newcommand{\dataset}{PhoneWatchUI}
\newcommand{\papertitle}{PhoneWatchUI: A Dataset of Smartphone and Smartwatch Images from App Stores}
\newcommand{\code}[1]{\texttt{#1}}

\definecolor{appleblue}{HTML}{2878A5}
\definecolor{playorange}{HTML}{BC6329}
\definecolor{othergray}{HTML}{BDBDBD}
\newcommand{\cellimg}[4]{\parbox[c][#2][c]{#1}{\centering{\setlength{\fboxsep}{0pt}\setlength{\fboxrule}{0.3pt}%
  \fcolorbox{black!25}{white}{\includegraphics[width=\dimexpr#1-0.6pt\relax,height=\dimexpr#3-0.6pt\relax,keepaspectratio]{#4}}}}}
\newcommand{\storerow}[3]{\begin{tikzpicture}[baseline=(b.center)]
  \node[minimum height=#2,minimum width=3.2mm,inner sep=0pt] (b) {\rotatebox{90}{\scriptsize\bfseries #3}};
  \draw[#1,line width=1pt] ([xshift=0.9mm]b.north east) -- ([xshift=0.9mm]b.south east);
\end{tikzpicture}\hspace{0.9mm}}
\newcommand{\gridhead}[2]{\parbox[b]{#1}{\centering\scriptsize\bfseries #2}}
\newlength{\gridcw}\newlength{\gridch}

\hypersetup{pdftitle={\papertitle},pdfauthor={Jialiang Wei and Walid Maalej}}

\begin{document}
\title{\papertitle}
\author{\IEEEauthorblockN{Jialiang Wei}
\IEEEauthorblockA{\textit{Hasso Plattner Institute}\\
\textit{University of Potsdam}\\ Potsdam, Germany\\ jialiang.wei@hpi.de}
\and
\IEEEauthorblockN{Walid Maalej}
\IEEEauthorblockA{\textit{Hasso Plattner Institute}\\
\textit{University of Potsdam}\\ Potsdam, Germany\\ walid.maalej@hpi.de}}
\maketitle

\begin{abstract}
Smartwatch apps are unique as they must condense functionality into a tiny screen of a few square centimeters. 
Yet, public GUI datasets are almost exclusively phone- or web-centric.
We present \dataset{}: 966.2k publisher-curated gallery image records from 72k store-specific app IDs on the US Apple App Store and Google Play, collected during October 2026.
The Apple App Store part covers 37.6k apps with explicit Apple Watch platform metadata: 143.8k watch and 212.2k phone images labeled by their source galleries.
The Google Play part covers 34.3k keyword-selected apps and 610.2k gallery images, provisionally labeled as watch, phone, or others by aspect ratio.
Image records link to app metadata (title, description, category, ratings, developer, price) and retain source URLs, device-label provenance, and dimensions where available.
We describe the dataset composition, image properties, and limitations.
\dataset{} enables research on cross-device GUI adaptation, wearable GUI retrieval and generation, as well as app-store analysis that links textual claims to visual evidence.
\end{abstract}

\begin{IEEEkeywords}
Smartwatch, Wearable Apps, App Store Mining, UI Design, GUI Dataset, watchOS, Wear OS, Apple Watch, Google Play, AI-Inspired UI Design
\end{IEEEkeywords}

\section{Introduction}
\begin{figure}[!t]
\centering
\setlength{\tabcolsep}{1pt}
\setlength{\gridcw}{0.243\columnwidth}\setlength{\gridch}{37mm}
\begin{tabular}{@{}cccc@{}}
\multicolumn{2}{c}{\parbox[b]{\dimexpr2\gridcw+1pt\relax}{\centering\scriptsize\bfseries Apple App Store\\[-1pt]\textcolor{appleblue}{\rule{\linewidth}{1pt}}}} &
\multicolumn{2}{c}{\parbox[b]{\dimexpr2\gridcw+1pt\relax}{\centering\scriptsize\bfseries Google Play\\[-1pt]\textcolor{playorange}{\rule{\linewidth}{1pt}}}} \\[2pt]
\cellimg{\gridcw}{\gridch}{\gridch}{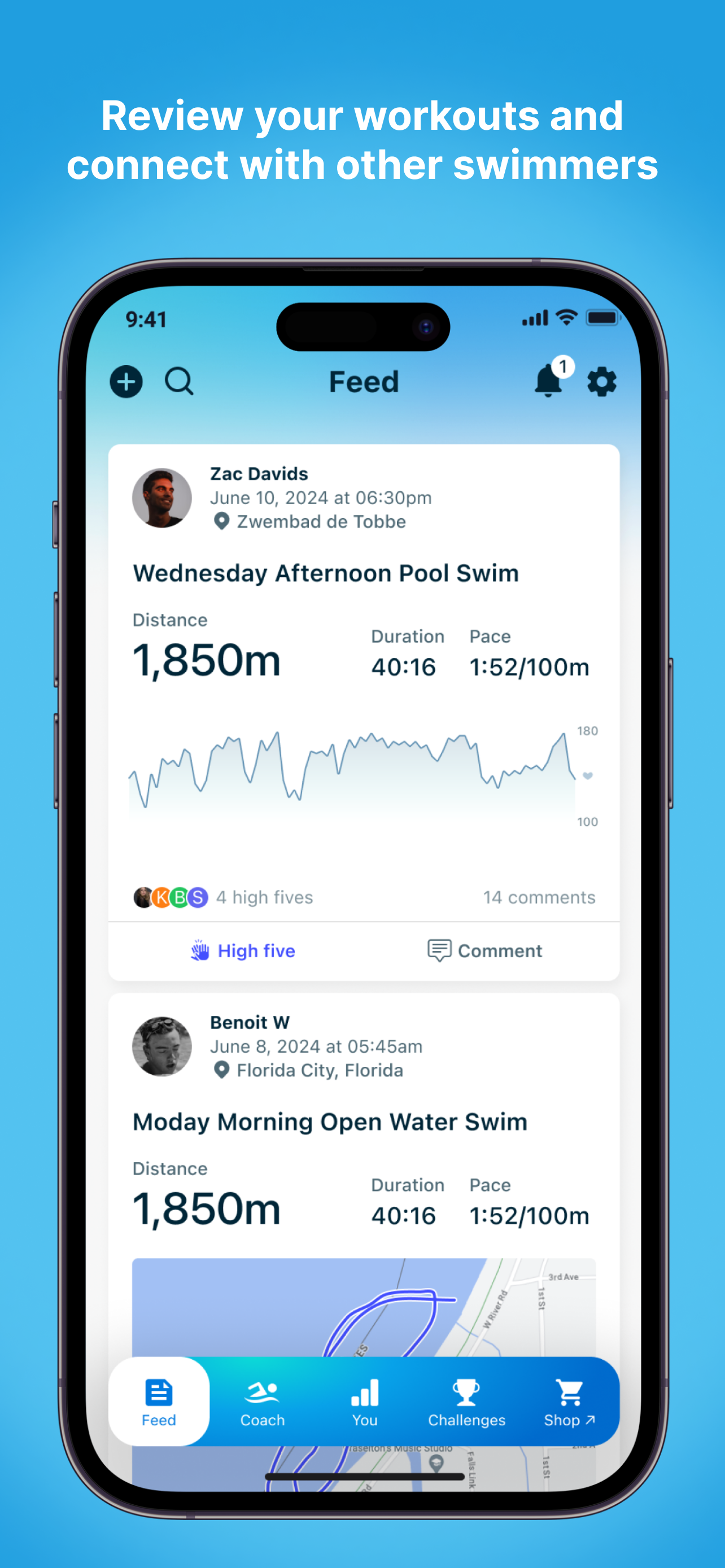} &
\cellimg{\gridcw}{\gridch}{21mm}{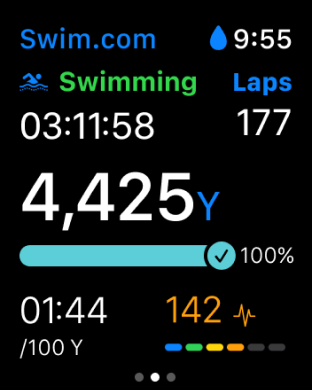} &
\cellimg{\gridcw}{\gridch}{\gridch}{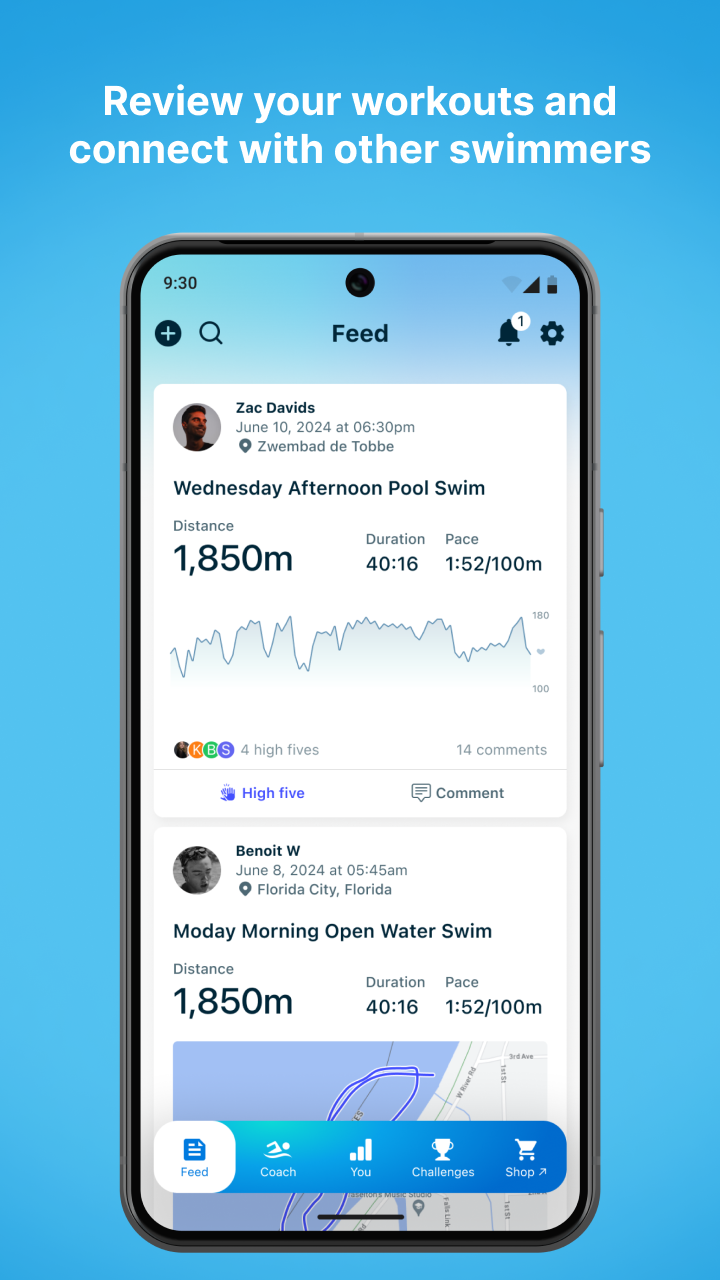} &
\cellimg{\gridcw}{\gridch}{21mm}{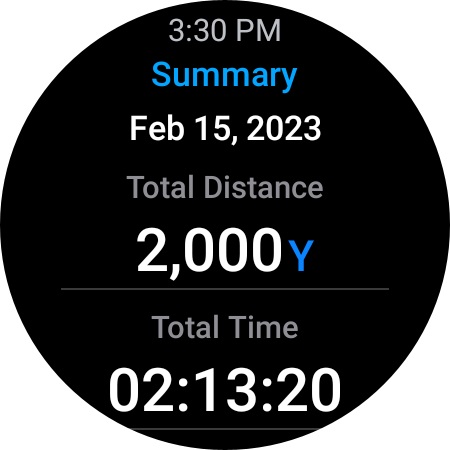} \\[2pt]
\scriptsize (a) Phone & \scriptsize (b) Watch & \scriptsize (c) Phone & \scriptsize (d) Watch
\end{tabular}
\caption{Swim.com phone and watch gallery images.}
\label{fig:cross-device}
\end{figure}

Smartwatches extend mobile computing to the wrist through two major ecosystems, Apple's watchOS and Google's Wear OS.
Designing for the wrist is not a matter of shrinking a phone screen: displays are tiny, round, or rounded, interactions last seconds, and the platforms add their own surfaces such as watch faces, complications, and tiles.
Developers adapt phone features to companion watch apps largely by hand, motivating empirically grounded guidance~\cite{Zhou:SurveyPatternsAdapting:2020,Chen:ComparativeStudySmartphone:2021}.
Figure~\ref{fig:cross-device} illustrates Swim.com across both stores and devices: the phone images present a workout feed, while the watch images focus on a few swimming metrics.

Data-driven GUI research, including design search~\cite{Feng:GalleryDCAutocreated:2022,Huang:SwireSketchbasedUser:2019,Chen:WireframebasedUIDesign:2020,Kolthoff:DatadrivenPrototypingNaturallanguagebased:2023,Wei:GUingMobileGUI:2025}, GUI prototyping and generation~\cite{Moran:MachineLearningBasedPrototyping:2020,Wei:BoostingGUIPrototyping:2023,Wei:AIInspiredUserInterface:2025}, and GUI understanding~\cite{Wang:Screen2WordsAutomaticMobile:2021,Liu:LearningDesignSemantics:2018}, depends on large screen collections such as Rico~\cite{Deka:RicoMobileApp:2017} and MUD~\cite{Feng:MUDLargeScaleNoiseFiltered:2024}.
These collections almost exclusively contain phone screens.
The only public dataset that pairs phone and watch UIs that we are aware of contains 310 watch captures from 101 Wear OS apps~\cite{Zhou:SurveyPatternsAdapting:2020}.
Software engineering research on wearable apps has mined reviews, permissions, and binaries~\cite{chauhan2016smartwatch,mujahid2018complaints,mujahid2018permission,Chen:ComparativeStudySmartphone:2021}, but not the apps' visual design at scale.

App stores offer a scalable alternative source.
App pages provide galleries that developers curate to present core features, alongside textual descriptions and market metadata~\cite{Harman:AppStoreMining:2012,Martin:SurveyAppStore:2017,Wei:GUingMobileGUI:2025}.
Crucially, the stores maintain device-specific information: the Apple App Store shows separate Apple Watch and iPhone galleries, and Google Play requires square Wear OS screenshots of at least $384{\times}384$ pixels~\cite{google2021wearscreenshots}.
In prior work, we showed that Google Play gallery images can train effective GUI retrieval and generation models for phone apps~\cite{Wei:GUingMobileGUI:2025,Wei:AIInspiredUserInterface:2025}.

This paper presents \dataset{}, an app store dataset of watch and phone images for wearable apps.
Our contribution includes:
\begin{itemize}
\item 966,186 image records from 71,928 store-specific app IDs, each linked to its app metadata, with same-app links between watch and phone galleries within each store where both are available;

\item scripts for app metadata and image collection, filtering, and verification that preserve gallery positions, source URLs, and available image hashes; and

\item a characterization of app categories, popularity, last update date, image properties, and research opportunities.
\end{itemize}

\section{Related Work}
\label{sec:related-work}

\textbf{GUI datasets.}
Table~\ref{tab:related} compares \dataset{} with representative GUI datasets.
Rico~\cite{Deka:RicoMobileApp:2017} collected 72k screens with view hierarchies by exploring 9.8k Android apps, and spawned semantic annotations~\cite{Liu:LearningDesignSemantics:2018}, screen summaries (Screen2Words~\cite{Wang:Screen2WordsAutomaticMobile:2021}), denoised layouts (Clay~\cite{Li:LearningDenoiseRaw:2022}), and topic labels (Enrico~\cite{Leiva:EnricoDatasetTopic:2020}).
MUD~\cite{Feng:MUDLargeScaleNoiseFiltered:2024} mined 18k noise-filtered UIs of 3.3k recent Android apps via LLM-guided exploration.
Further datasets cover design search (VINS~\cite{Bunian:VINSVisualSearch:2021}), web pages (WebUI~\cite{Wu:WebUIDatasetEnhancing:2023}), device control (Android in the Wild~\cite{Rawles:AndroidWildLargescale:2023}), recent apps (MobileViews~\cite{Gao:MobileViewsMillionscaleDiverse:2025}), and phone--tablet pairs (Papt~\cite{Hu:PairwiseDatasetGUI:2023}).
Closest to our source, ScreenRepo~\cite{Wei:GUingMobileGUI:2025} contains 303k phone screenshots extracted from Google Play gallery images; its captioned subset, SCapRepo, contains 135k screenshots.
None of these datasets targets smartwatches.
Zhou et al.~\cite{Zhou:SurveyPatternsAdapting:2020} manually mapped phone to watch UIs of 101 popular Wear OS apps and derived adaptation patterns, a valuable but small sample of a single platform.

\textbf{Wearable apps.}
Chauhan et al.~\cite{chauhan2016smartwatch} characterized about 14k early smartwatch apps by price, developer, and category.
Mujahid et al.\ studied user complaints~\cite{mujahid2018complaints} and permission issues~\cite{mujahid2018permission} of Android wearable apps, Liu and Lin characterized Wear OS itself~\cite{liu2016wearos}, and Chen et al.\ compared phone and watch versions of Android apps~\cite{Chen:ComparativeStudySmartphone:2021}.

\begin{table}[t]
\centering
\caption{GUI dataset comparison (rounded counts; --: unreported).}
\label{tab:related}
\footnotesize
\setlength{\tabcolsep}{2.6pt}
\begin{tabular}{@{}llrrl@{}}
\toprule
Dataset & Devices & Images & Apps & Source \\
\midrule
Rico~\cite{Deka:RicoMobileApp:2017} & Phone & 72k & 9.8k & App exploration \\
MUD~\cite{Feng:MUDLargeScaleNoiseFiltered:2024} & Phone & 18k & 3.3k & LLM-guided crawl \\
ScreenRepo~\cite{Wei:GUingMobileGUI:2025} & Phone & 303k & -- & Google Play gallery \\
Zhou et al.~\cite{Zhou:SurveyPatternsAdapting:2020} & Phone, watch & 1.0k & 101 & UI crawl + labeling \\
\midrule
\shortstack[l]{\dataset{}\\(Apple App Store)} & Watch, phone & 356.0k & 37.6k & \shortstack[l]{Apple App Store \\ gallery} \\
\shortstack[l]{\dataset{}\\(Google Play)} & Watch, phone & 610.2k & 34.3k & Google Play gallery \\
\bottomrule
\end{tabular}
\end{table}

\section{Dataset Construction}
\label{sec:construction}
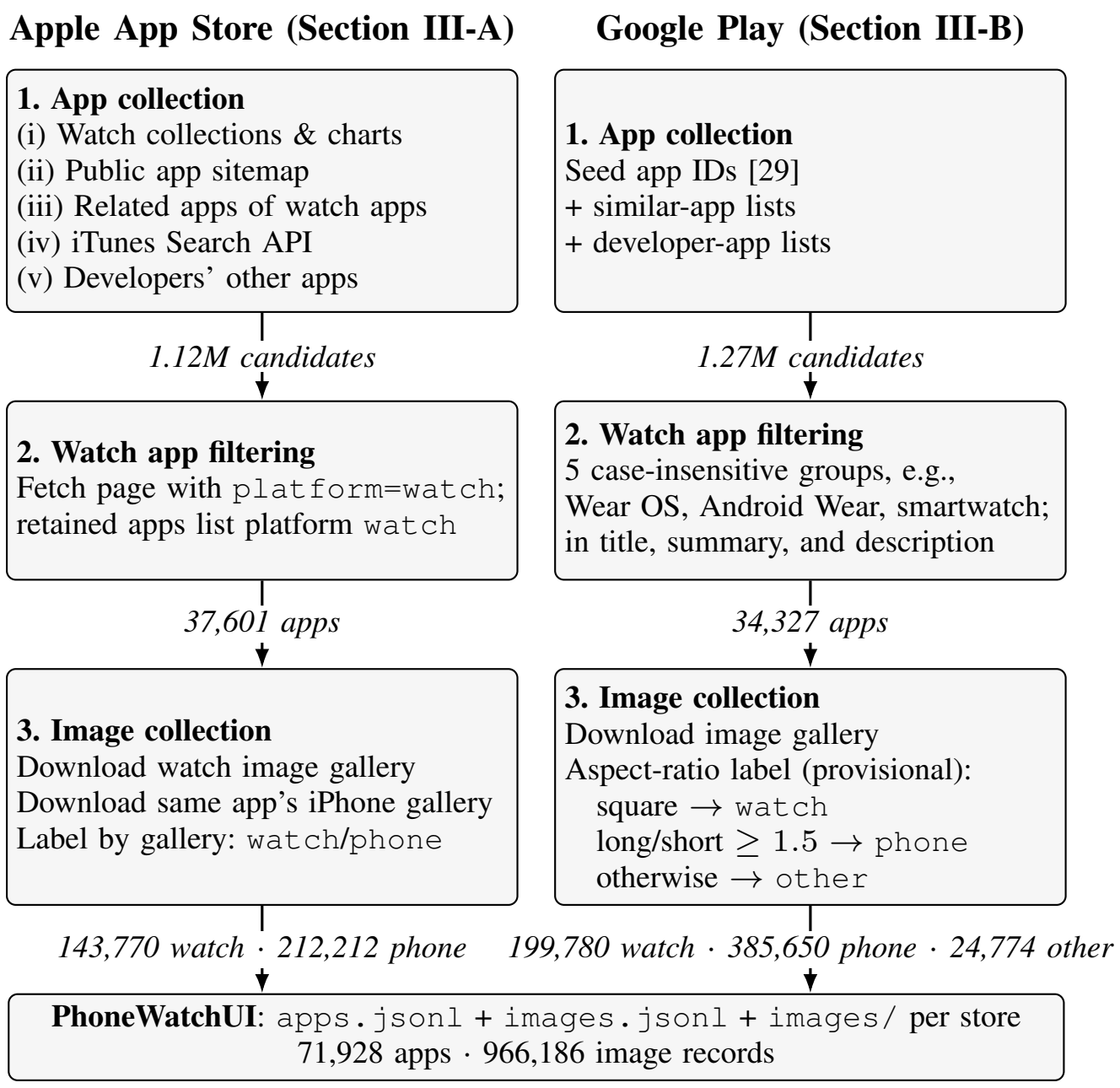
\begin{figure}[t]
\centering
\begin{tikzpicture}[
  font=\scriptsize,
  lane/.style={draw, rounded corners=2pt, fill=gray!7, align=left, text width=38.5mm, inner sep=2.2pt, anchor=north},
  head/.style={font=\footnotesize\bfseries, align=center},
  outbox/.style={draw, rounded corners=2pt, fill=gray!7, align=center, text width=81mm, inner sep=2.5pt},
  cnt/.style={fill=white, inner sep=1pt, font=\scriptsize\itshape, align=center},
  arr/.style={-{Latex[length=1.5mm]}, semithick}]
\node[head] (ha) at (0,0) {Apple App Store (Section \ref{sec:apple})};
\node[head] (hg) at (43mm,0) {Google Play (Section \ref{sec:google})};
\node[lane, minimum height=19mm] (a1) at (0,-3mm) {\textbf{1. App collection}\\
  (i)~Watch collections \& charts\\
  (ii)~Public app sitemap\\
  (iii)~Related apps of watch apps\\
  (iv)~iTunes Search API\\
  (v)~Developers' other apps};
\node[lane, minimum height=19mm] (g1) at (43mm,-3mm) {\textbf{1. App collection}\\
  Seed app IDs~\cite{Wei:GettingInspirationFeature:2024}\\
  + similar-app lists\\
  + developer-app lists};
\node[lane, minimum height=14mm] (a2) at (0,-29mm) {\textbf{2. Watch app filtering}\\
  Fetch page with \code{platform=watch};\\
  retained apps list platform \code{watch}};
\node[lane, minimum height=14mm] (g2) at (43mm,-29mm) {\textbf{2. Watch app filtering}\\
  5 case-insensitive groups, e.g.,\\
  Wear OS, Android Wear, smartwatch;\\
  in title, summary, and description};
\node[lane, minimum height=18.5mm] (a3) at (0,-50mm) {\textbf{3. Image collection}\\
  Download watch image gallery\\
  Download same app's iPhone gallery\\
  Label by gallery: \code{watch}/\code{phone}};
\node[lane, minimum height=18.5mm] (g3) at (43mm,-50mm) {\textbf{3. Image collection}\\
  Download image gallery\\
  Aspect-ratio label (provisional):\\
  \quad square $\rightarrow$ \code{watch}\\
  \quad long/short $\geq 1.5$ $\rightarrow$ \code{phone}\\
  \quad otherwise $\rightarrow$ \code{other}};
\node[outbox, anchor=north] (o) at (21.5mm,-75.5mm) {\textbf{\dataset{}}: \code{apps.jsonl} + \code{images.jsonl} + \code{images/} per store\\
  71,928 apps $\cdot$ 966,186 image records};
\draw[arr] (a1) -- node[cnt]{1.12M candidates} (a2);
\draw[arr] (g1) -- node[cnt]{1.27M candidates} (g2);
\draw[arr] (a2) -- node[cnt]{37,601 apps} (a3);
\draw[arr] (g2) -- node[cnt]{34,327 apps} (g3);
\draw[arr] (a3.south) -- node[cnt]{143,770 watch $\cdot$ 212,212 phone} (a3.south |- o.north);
\draw[arr] (g3.south) -- node[cnt]{199,780 watch $\cdot$ 385,650 phone $\cdot$ 24,774 other} (g3.south |- o.north);
\end{tikzpicture}
\caption{Data collection for Apple App Store and Google Play.}
\label{fig:pipeline}
\end{figure}

We describe the workflow to collect images from the Apple App Store and Google Play, as depicted in Figure~\ref{fig:pipeline}.
This collection was performed during October 2026.

\subsection{Apple App Store}
\label{sec:apple}
\subsubsection{App collection}
The Apple App Store does not provide a complete catalog of watch apps. We therefore combined five complementary sources to discover as many candidate apps as possible:
(i)~official US Apple Watch collections and category charts, including embedded page data;
(ii)~all pages in the public app sitemap;
(iii)~related apps recommended on confirmed watch-app pages;
(iv)~the iTunes Search API, using queries combining watch terms and use cases, supplemented with terms from confirmed watch-app names; and
(v)~other apps published by developers of confirmed watch apps.
This step collected 1.12 million candidate apps.

\subsubsection{Watch app filtering}
For each candidate app, we requested its Apple App Store page with \code{platform=watch} and parsed the embedded metadata.
All 37,601 retained apps explicitly included \code{watch} in their platform metadata.

\subsubsection{Image collection}
We attempted to download every image position in each app's Apple Watch gallery at the resolution provided by the Apple App Store.
We checked each file's format, dimensions, and completeness.
We then verified that each iPhone page belonged to the same app and downloaded its gallery in the same way, excluding iPad, Mac, and other device media.
Each image was labeled \code{watch} or \code{phone} according to its source gallery.

\subsection{Google Play}
\label{sec:google}
\subsubsection{App collection}
We started from a seed list of package IDs from an earlier Google Play corpus of Wei et al.~\cite{Wei:GettingInspirationFeature:2024} and re-fetched metadata for every app from the US/English store, expanding the candidate set through the apps' similar-app and developer-app lists.
For each app we collected app metadata, including description, category, ratings, price, developer, permissions, data-safety section, and image URLs.
This step collected 1.27 million candidate apps.

\subsubsection{Watch app filtering}
Unlike Apple App Store, apps on Google Play does not expose device information, so we filter the watch-related apps by matching the title, summary, and description against five case-insensitive groups: (1)~Wear OS/WearOS; (2)~Android Wear; (3)~smartwatch; (4)~watch face; and (5)~watch app, watch application, or watch version.
The rules also cover spacing, hyphenation, and plural variants.
We retained 34,327 apps matching any group.

\subsubsection{Image collection}
We downloaded Google Play gallery images at original resolution.
Because the parsed gallery does not attribute images to devices, we assign a device label from the aspect ratio.
Square images are labeled \code{watch}, motivated by Google Play's 1:1 requirement for Wear OS screenshots~\cite{google2021wearscreenshots}.
Our manual inspection on 1,000 randomly sampled square images suggests that more than 99\% of them depict watch-app interfaces or watch faces.
Following GUing~\cite{Wei:GUingMobileGUI:2025}, images whose long-to-short side ratio is at least 1.5 are labeled \code{phone}.
All remaining images are labeled \code{other}.
Every Google Play label records this rule and its threshold.

\begin{table}[t]
\centering
\caption{Main fields of the two record types (both stores).}
\label{tab:schema}
\footnotesize
\setlength{\tabcolsep}{3pt}
\begin{tabular}{@{}>{\raggedright\arraybackslash}p{35mm}>{\raggedright\arraybackslash}p{47mm}@{}}
\toprule
Field & Description \\
\midrule
\multicolumn{2}{@{}l}{\textit{\code{apps.jsonl} (one row per app)}} \\
\code{app\_id}, \code{store} & Apple App Store numeric ID or Android package name \\
\code{title}, \code{description} & App title and description as observed \\
\code{source\_url} & US/English app URL \\
\code{metadata} & App attributes: developer, category, price, rating, version, update date, \dots \\
\midrule
\multicolumn{2}{@{}l}{\textit{\code{images.jsonl} (one row per gallery position)}} \\
\code{image\_id}, \code{app\_id} & Stable position identity; app join key within a store \\
\code{gallery}, \code{position} & Source gallery and zero-based ordinal \\
\code{image\_path} & Relative file path \\
\code{width}, \code{height}, \code{aspect\_ratio} & Pixel dimensions and width/height ratio \\
\code{device} & \code{watch}, \code{phone}, or \code{other} \\
\code{device\_label\_source} & Label method and ratio threshold \\
\code{source\_url}, \code{sha256} & Original image URL and SHA-256 hash \\
\bottomrule
\end{tabular}
\end{table}

\section{Dataset Characteristics}
\label{sec:characteristics}
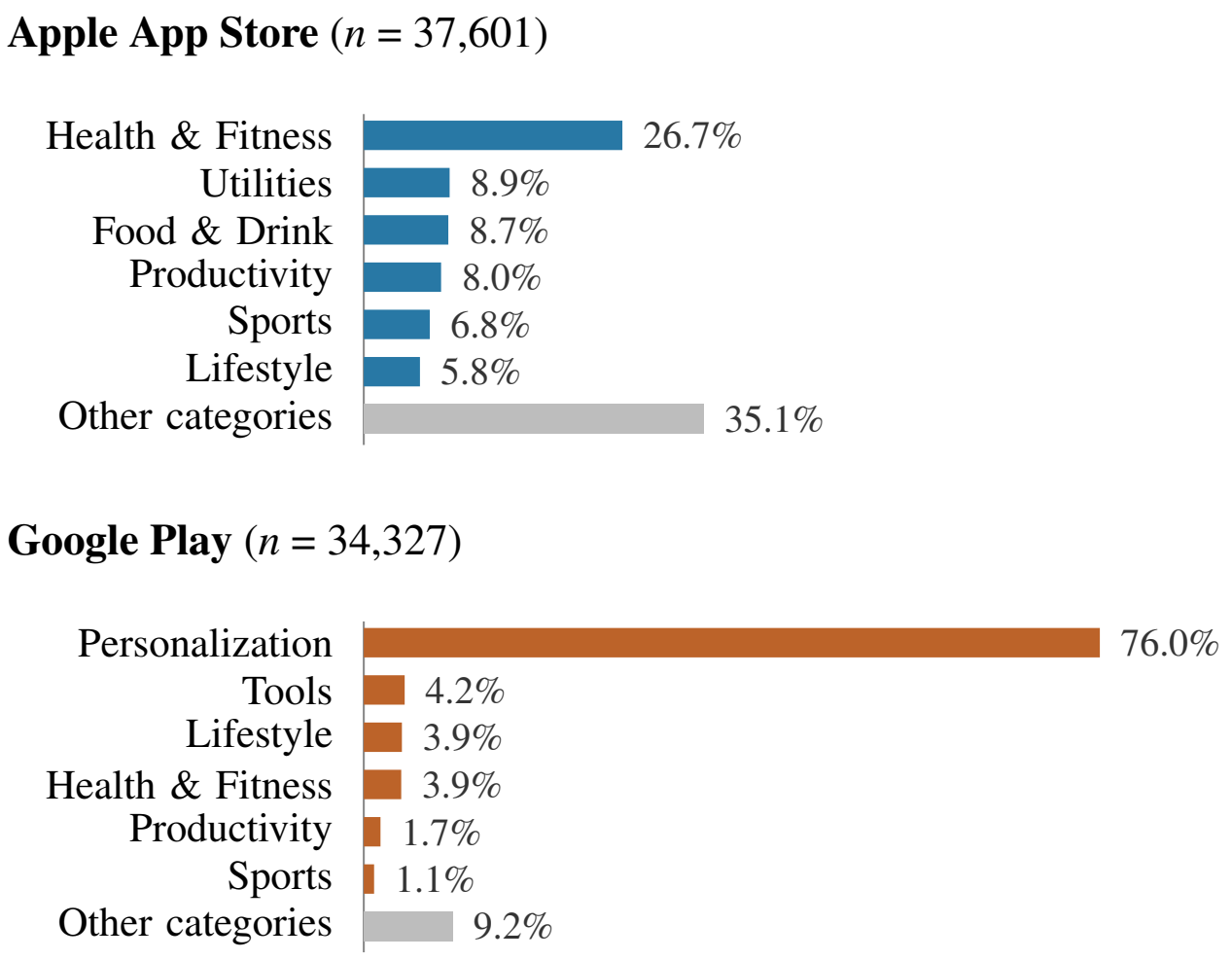
\begin{figure}[t]
\centering
\begin{tikzpicture}
\begin{groupplot}[
  group style={group size=1 by 2,vertical sep=12mm},
  width=56mm,height=22.5mm,scale only axis,
  xmin=0,xmax=85,ymin=-0.45,ymax=6.55,y dir=reverse,
  ytick={0,1,2,3,4,5,6},
  xbar,/pgf/bar width=2.0mm,/pgf/bar shift=0pt,
  hide x axis,
  axis y line*=left,y axis line style={black!45,line width=0.4pt},
  ytick style={draw=none},
  tick label style={font=\fontsize{8}{9}\selectfont},
  yticklabel style={text width=24mm,align=right,inner sep=1.5pt},
  title style={at={(0,1)},anchor=south west,xshift=-25.5mm,yshift=0.6mm,
    font=\fontsize{8.5}{10}\selectfont},
  point meta=x,
  nodes near coords={\pgfmathprintnumber[fixed,precision=1,zerofill,assume math mode=true]{\pgfplotspointmeta}\%},
  every node near coord/.append style={font=\fontsize{7.5}{9}\selectfont,anchor=west,xshift=0.5pt,text=black!80},
  clip=false,enlarge x limits=false,
]
\nextgroupplot[title={\textbf{Apple App Store} (\textit{n}~=~37,601)},yticklabels={{Health \& Fitness},{Utilities},{Food \& Drink},{Productivity},{Sports},{Lifestyle},{Other categories}}]
\addplot[fill=appleblue,draw=none] coordinates {(26.70407702,0) (8.86678546,1) (8.73115077,2) (7.98383022,3) (6.81630808,4) (5.80303715,5)};
\addplot[fill=othergray,draw=none] coordinates {(35.09481131,6)};
\nextgroupplot[title={\textbf{Google Play} (\textit{n}~=~34,327)},yticklabels={{Personalization},{Tools},{Lifestyle},{Health \& Fitness},{Productivity},{Sports},{Other categories}}]
\addplot[fill=playorange,draw=none] coordinates {(76.00722463,0) (4.22408017,1) (3.93859061,2) (3.86867480,3) (1.72750313,4) (1.06621610,5)};
\addplot[fill=othergray,draw=none] coordinates {(9.16771055,6)};
\end{groupplot}
\end{tikzpicture}
\caption{App shares by primary app category in each store, on a common scale. The six largest categories are shown separately; ``Other categories'' combines the rest.}
\label{fig:categories}
\end{figure}

Tables~\ref{tab:schema} and~\ref{tab:stats} summarize the schema and statistics; Figs.~\ref{fig:phone-examples} and~\ref{fig:watch-examples} show phone and watch examples.
The images occupy 822.6 GB (304.1 GB Apple App Store, 518.6 GB Google Play) and the JSONL metadata 946 MB.
All gallery positions and duplicates are retained, with source URLs, device-label provenance, and hashes for downloaded files.
App counts use store-specific IDs, so a product can appear in both stores. The 62,151 IDs with both device labels link galleries at the app level, not individual screens.
App percentages use all IDs in each store; missing rating counts are not zero. We describe the stores separately because their composition differs (Fig.~\ref{fig:categories}).

\begin{table}[t]
\centering
\caption{Dataset statistics. Per-app values are median (mean) over apps with that device label.}
\label{tab:stats}
\footnotesize
\setlength{\tabcolsep}{3pt}
\begin{tabular}{@{}lrr@{}}
\toprule
 & Apple App Store & Google Play \\
\midrule
Apps & 37,601 & 34,327 \\
Known developer IDs & 24,401 & 4,875 \\
Primary category labels & 42 & 48 \\
Free apps & 88.2\% & 30.1\% \\
\midrule
Downloaded files & 355,979 & 610,201 \\
Image records & 355,982 & 610,204 \\
\quad watch & 143,770 & 199,780 \\
\quad phone & 212,212 & 385,650 \\
\quad other & 0 & 24,774 \\
Device label & store gallery & aspect ratio (prov.) \\
Apps with watch images & 37,590 & 30,374 \\
Apps with watch \& phone images & 35,266 & 26,885 \\
Watch images per app & 4 (3.8) & 6 (6.6) \\
Phone images per app & 6 (6.0) & 9 (12.6) \\
\bottomrule
\end{tabular}
\end{table}

\begin{figure}[t]
\centering
\setlength{\tabcolsep}{1.2pt}
\setlength{\gridcw}{0.292\columnwidth}\setlength{\gridch}{29mm}
\begin{tabular}{@{}cccc@{}}
 & \gridhead{\gridcw}{Screenshot} & \gridhead{\gridcw}{Surrounded screenshot} & \gridhead{\gridcw}{Promotion} \\[2pt]
\storerow{appleblue}{\gridch}{Apple App Store} &
\cellimg{\gridcw}{\gridch}{\gridch}{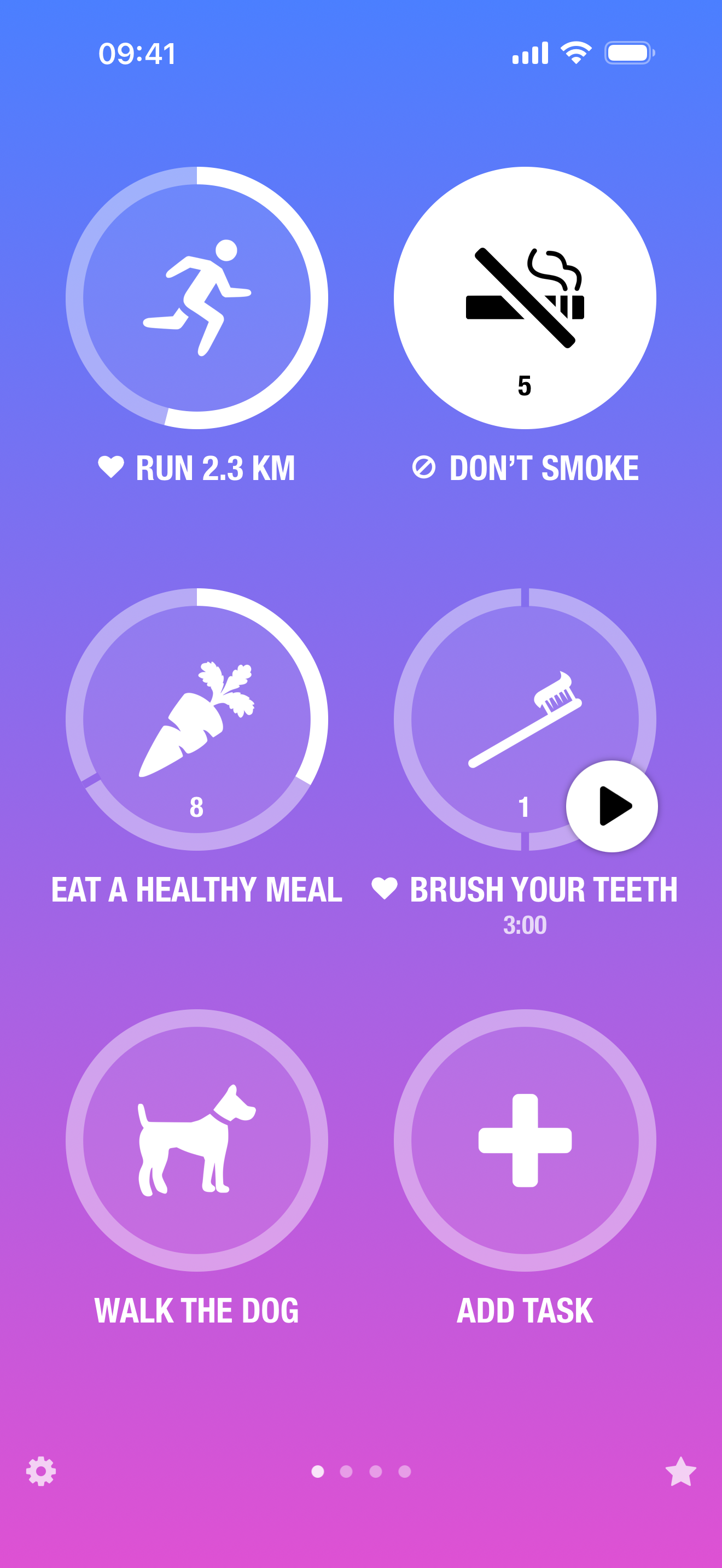} &
\cellimg{\gridcw}{\gridch}{\gridch}{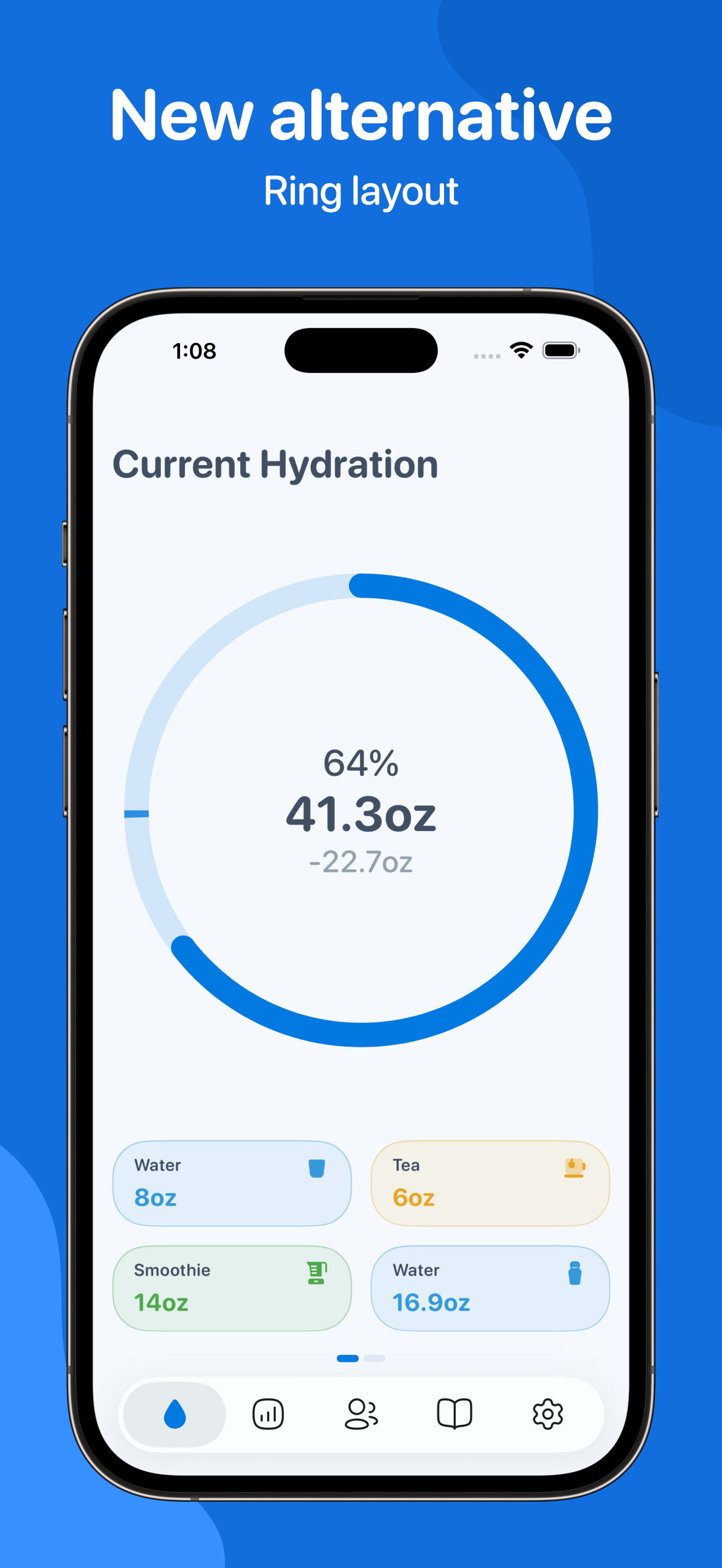} &
\cellimg{\gridcw}{\gridch}{\gridch}{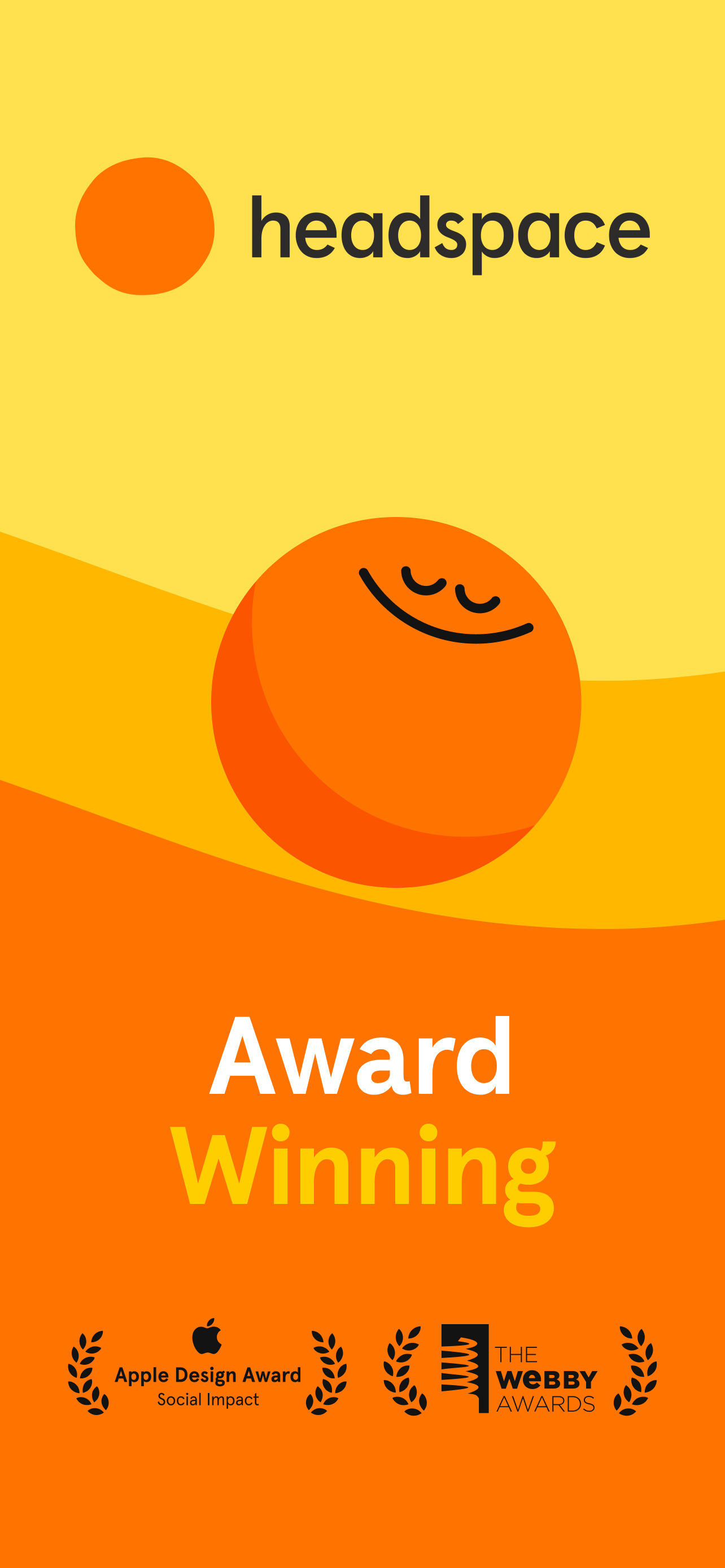} \\[3pt]
\storerow{playorange}{\gridch}{Google Play} &
\cellimg{\gridcw}{\gridch}{\gridch}{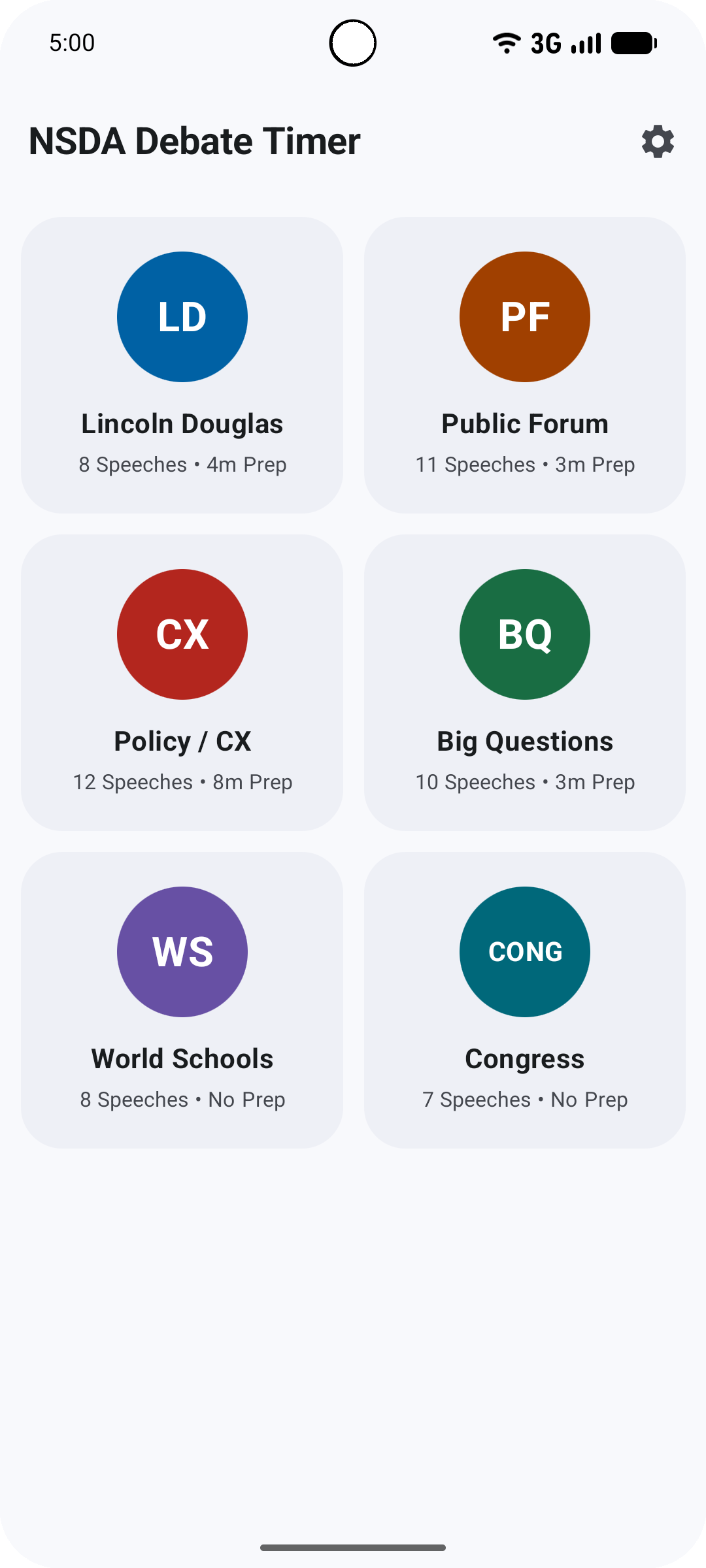} &
\cellimg{\gridcw}{\gridch}{\gridch}{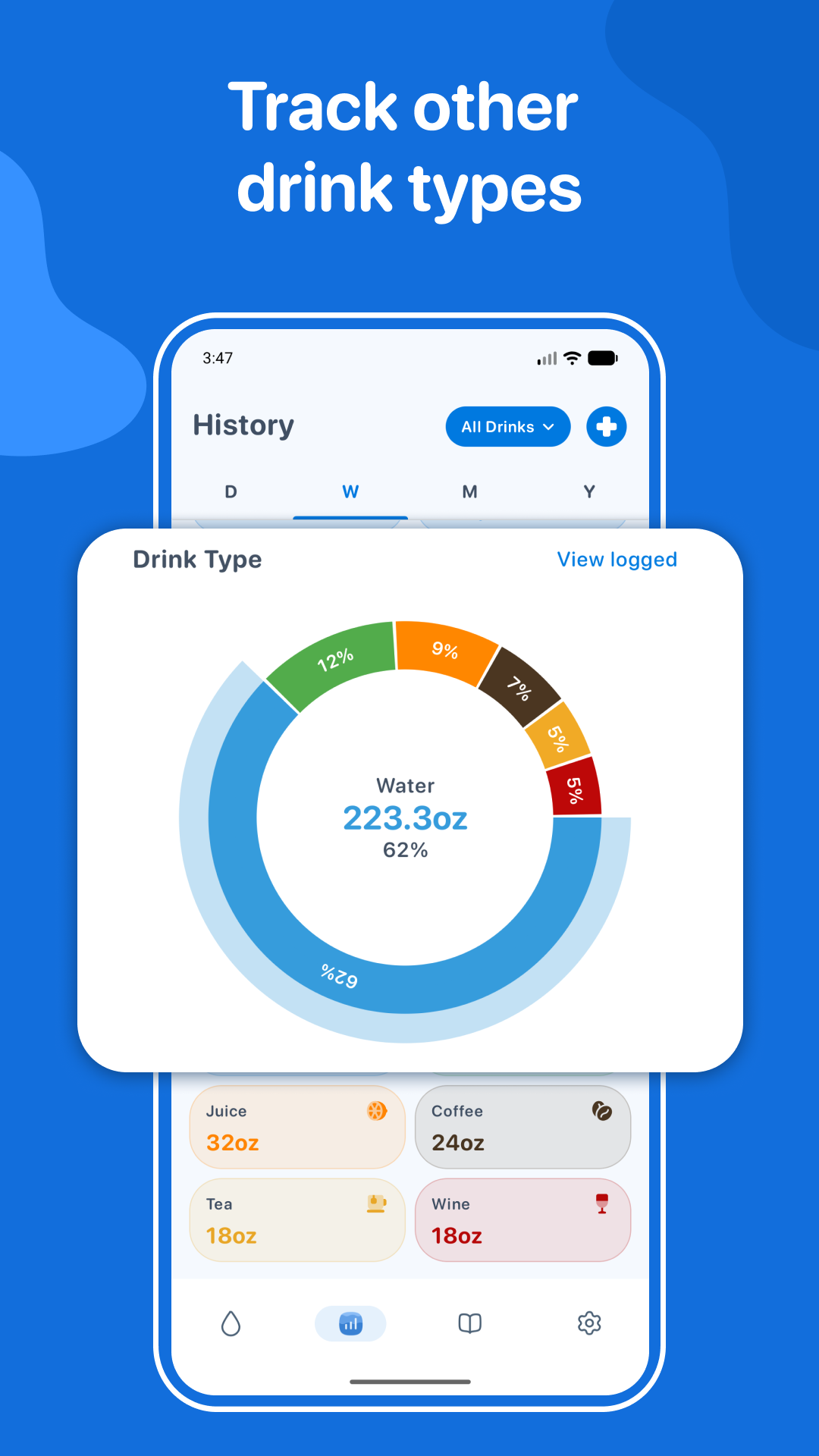} &
\cellimg{\gridcw}{\gridch}{\gridch}{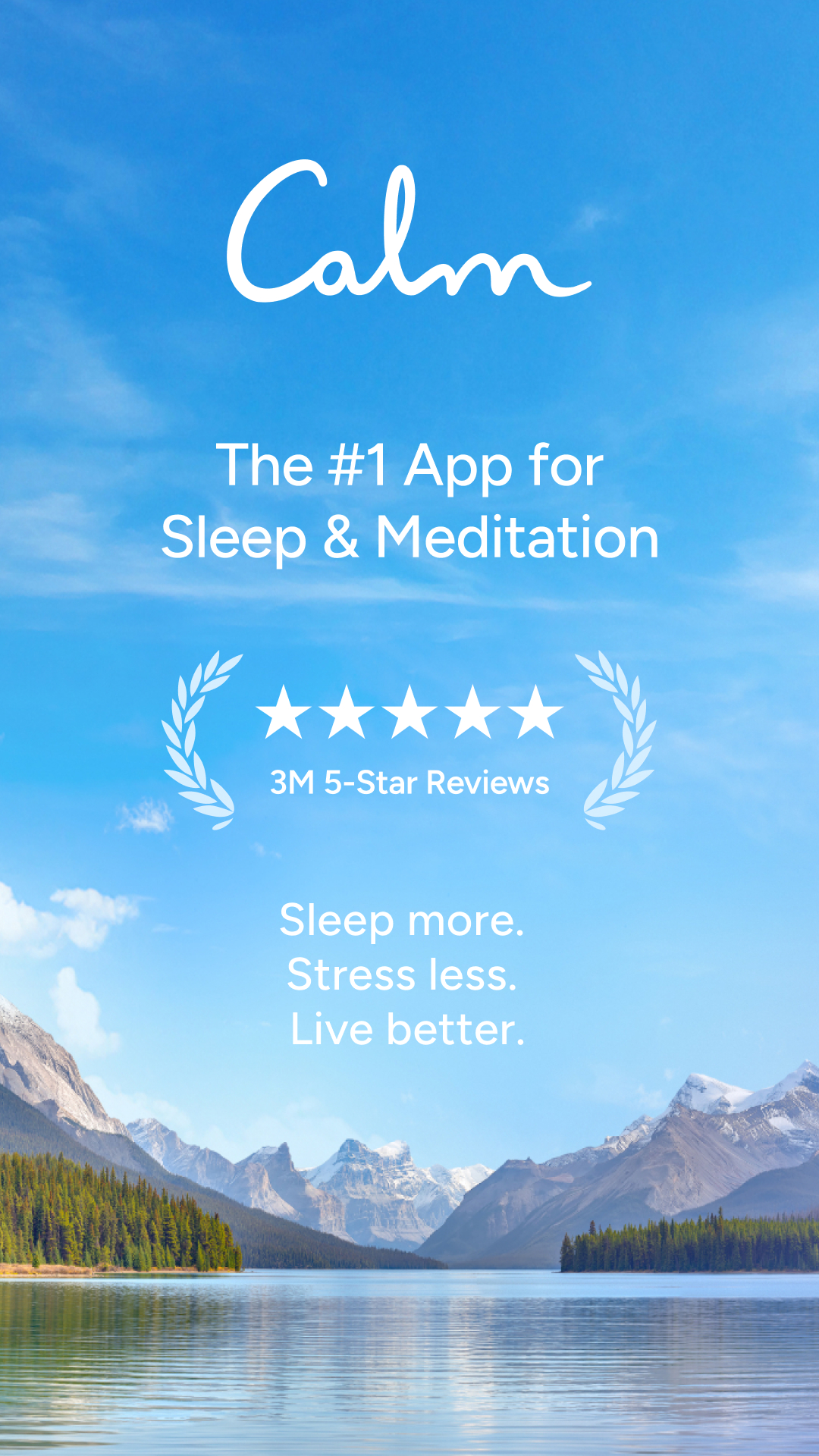}
\end{tabular}
\caption{Phone-labeled examples: screenshots, surrounded screenshots, and promotional images, following GUing~\cite{Wei:GUingMobileGUI:2025}.}
\label{fig:phone-examples}
\end{figure}

\begin{figure}[t]
\centering
\setlength{\tabcolsep}{1.2pt}
\setlength{\gridcw}{0.292\columnwidth}\setlength{\gridch}{24mm}
\begin{tabular}{@{}cccc@{}}
 & \gridhead{\gridcw}{App UI} & \gridhead{\gridcw}{Watch face} & \gridhead{\gridcw}{Promotion} \\[2pt]
\storerow{appleblue}{\gridch}{Apple App Store} &
\cellimg{\gridcw}{\gridch}{\gridch}{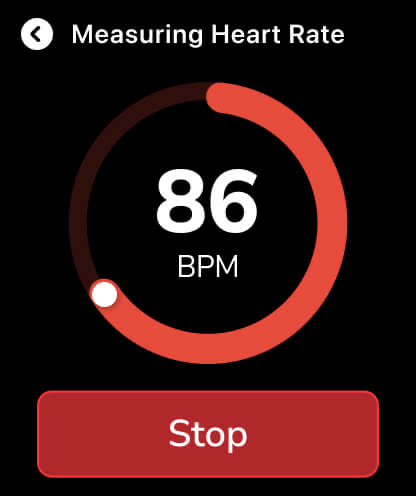} &
\cellimg{\gridcw}{\gridch}{\gridch}{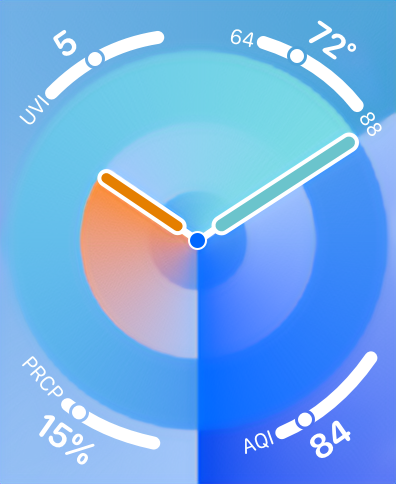} &
\cellimg{\gridcw}{\gridch}{\gridch}{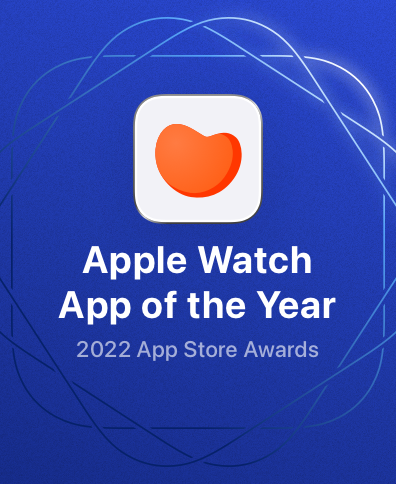} \\[3pt]
\storerow{playorange}{\gridch}{Google Play} &
\cellimg{\gridcw}{\gridch}{\gridch}{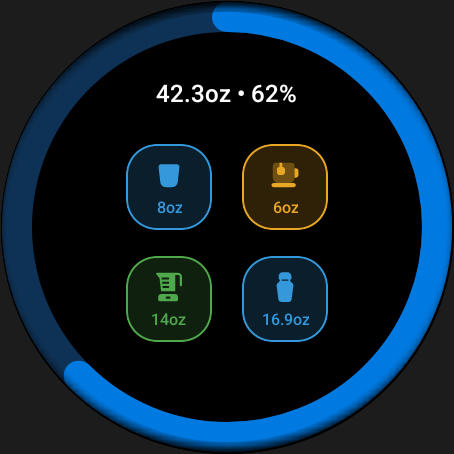} &
\cellimg{\gridcw}{\gridch}{\gridch}{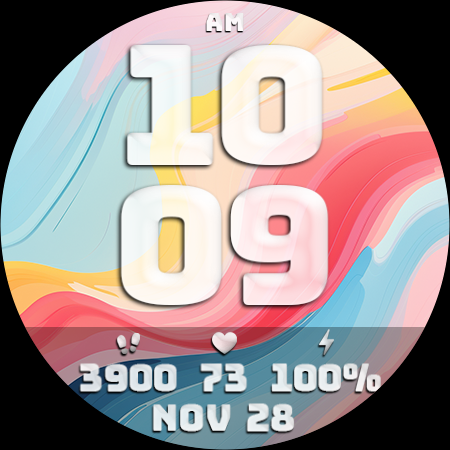} &
\cellimg{\gridcw}{\gridch}{\gridch}{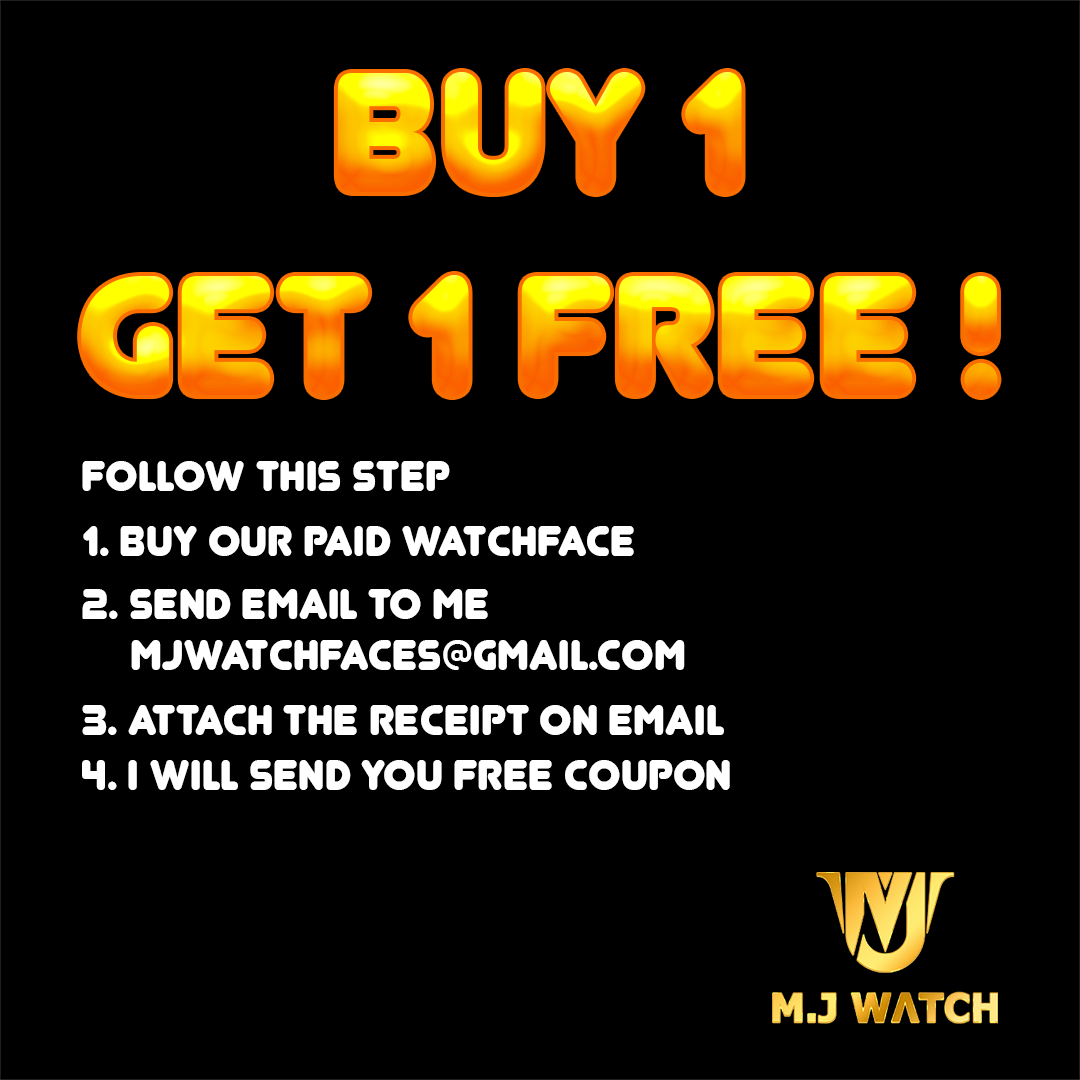}
\end{tabular}
\caption{Watch-labeled examples: app UIs, watch faces, and promotional images.}
\label{fig:watch-examples}
\end{figure}

\subsection{Apple App Store}
Health \& Fitness is the largest category (26.7\%), followed by Utilities (8.9\%), Food \& Drink (8.7\%), and Productivity (8.0\%).
The metadata identify 24,401 developers (one app lacks a developer ID), and 88.2\% of apps are free.
Positive rating counts are recorded for 54.6\% of apps, with a median of 9 among these apps; six apps have missing counts. The last recorded update falls in 2026 for 65.2\% of apps.

The galleries are small and standardized.
The medians are four watch and six phone images per app. All but 12 watch images use one of the six sizes permitted by the Apple App Store screenshot specification~\cite{apple2026screenshots}, most commonly $368{\times}448$ (26.7\%), $416{\times}496$ (20.5\%), and $312{\times}390$ (15.2\%).
Phone images are almost all portrait (98.6\%), led by $1242{\times}2688$ (33.8\%).
Watch and phone images are available for 35,266 apps (93.8\%), and all downloaded files are byte-distinct.

\subsection{Google Play}
Personalization, the category under which watch faces are listed, accounts for 76.0\% of the apps, and no other category exceeds 4.3\%.
The apps come from only 4,875 developers, are mostly paid (69.9\%), and are concentrated: the ten largest developers publish 14.1\% of the apps, the largest alone 813.
Positive rating counts are recorded for 20.5\% of apps, with a median of 98 among these apps; 27,284 apps have missing counts. The last recorded update falls in 2026 for 60.4\% of apps.

The galleries are larger than on the Apple App Store (a median of six watch and nine phone images) and more heterogeneous: square images alone span 462 resolutions, led by $450{\times}450$ (46.5\%) and $512{\times}512$ (19.0\%).
Watch and phone images are available for 26,885 apps (78.3\%).
Personalization apps contribute 87.5\% of the square images, indicating a strong bias toward watch face designs (Fig.~\ref{fig:watch-examples}, middle column); studies of functional watch-app UIs need additional content filtering.

Galleries repeat images: only 60.7\% of the files are byte-distinct, and 58.6\% of the gallery positions share their bytes with at least one other position, reflecting templated galleries and reused promotional assets.
For model evaluation, identical hashes should stay in the same data split to reduce train--test leakage.
Furthermore, 3,867 apps with images contain no square image, so a watch-related keyword does not guarantee a watch-shaped image.

\section{Research Opportunities}
\dataset{} enables research questions that existing phone-centric GUI datasets cannot address.

\textbf{RQ1: How do apps adapt their features from phone to watch?}
For the 62,151 apps with both watch and phone images, researchers can compare gallery features to study which are kept, reduced, or dropped on the watch and how this varies by category.
This would turn the manual patterns of Zhou et al.~\cite{Zhou:SurveyPatternsAdapting:2020} into store-scale empirical guidance.

\textbf{RQ2: Is the watch-face market driven by template-based mass publishing?}
The Google Play part is dominated by paid watch faces (76.0\% Personalization, 69.9\% paid) from a few prolific developers (the largest publishes 813 apps), and 58.6\% of gallery positions share their bytes with at least one other position.
Hashes, developer IDs, and prices enable studies of templating, clone detection, and pricing strategies.

\textbf{RQ3: Are watch claims in app stores backed by visual evidence?}
Linking descriptions, metadata, and galleries reveals whether watch-related claims come with watch images (3,867 Google Play apps with images contain no square image), how developers present watch apps, whether presentations follow store guidelines~\cite{google2021wearscreenshots,apple2026screenshots}, and whether presentation practices relate to ratings.

\textbf{RQ4: How well do vision-language models understand small-screen GUIs?}
The images and descriptions support benchmarks for device and content-type classification (app UI, watch face, promotion) and description-to-image retrieval, and provide training data for watch-specific GUI retrieval and generation~\cite{Wei:GUingMobileGUI:2025,Wei:BoostingGUIPrototyping:2023}.

\section{Limitations}
\textbf{Coverage.}
The dataset was collected during October 2026, and targets the US store in English.
Its coverage is bounded by the discovery sources and does not constitute a census.
Google Play apps with non-English descriptions or without the selected keywords are underrepresented.
Keyword matches alone do not establish watch compatibility, as they may refer to guides, companion tools, or future features.

\textbf{Device labels.}
Apple App Store device labels reflect gallery assignments.
Google Play labels are provisional.
Although more than 99\% of our 1,000 sampled square images are watch-related, the images labeled as phone by aspect ratio may include portrait tablet images.

\textbf{Content.}
The images are marketing assets rather than UI screenshots and lack view hierarchies.
Many include device mock-ups or captions, while some contain no app UI.
Phone images can be processed using GUing's image classifier and cropper~\cite{Wei:GUingMobileGUI:2025}, whereas comparable processing of watch images requires further work.
Repeated images must be accounted for when constructing data splits to avoid train--test leakage.

\textbf{Rights.}
Images remain the property of their owner and publishers. 
The dataset is collected and shared for research purpose only.

\textbf{Size.}
The full dataset requires about 823 GB of storage, which may limit its accessibility and use.

\section{Data Availability}
\label{sec:availability}
The app and image metadata are available at \url{https://huggingface.co/datasets/Jl-wei/PhoneWatchUI}, including source URLs for downloading the images and SHA-256 hashes for verifying their integrity.
Upon acceptance, we will archive a versioned release with a DOI on Zenodo.

\section{Conclusion}
\dataset{} provides 966.2k gallery image records from 72k store-specific app IDs on the Apple App Store and Google Play, covering phone, watch, and other images linked to app metadata, source URLs, and device labels.
We hope it will support research on watch app development, app store mining, and AI-inspired UI design.

\balance
\bibliographystyle{IEEEtran}
\bibliography{ref}
\end{document}